\documentclass[reprint,aip,jcp]{revtex4-1}
\usepackage{graphicx}
\usepackage{dcolumn}
\usepackage{bm}
\usepackage{subfigure}
\usepackage{tikz}
\usepackage{color}
\usepackage{lineno}
\usepackage{marginnote}
\usepackage{comment}

\usepackage{wasysym}
\usepackage{amssymb}	

\let\baraccent=\= 
\renewcommand{\=}[1]{\stackrel{#1}{=}} 

\usetikzlibrary{shapes.callouts}
\usetikzlibrary{shapes.geometric,shapes.arrows,decorations.pathmorphing}
\usetikzlibrary{matrix,chains,scopes,positioning,arrows,fit}

\linethickness{0.4mm}
\definecolor{mma1}{rgb}{0.3725,0.5098,0.7020}
\definecolor{mma2}{rgb}{0.8745,0.6078,0.2039}
\definecolor{mma3}{rgb}{0.507813,0.714844,0.2039}
\definecolor{mma4}{rgb}{0.9137,0.3882,0.2398}

\tikzset{
  laser/.style   = { ultra thick, mma4},
  connect/.style = { dashed, red },
  notice/.style  = { draw, rectangle callout, callout relative pointer={#1} },
  label/.style   = { text width=2cm },
  laser1/.style   = { ultra thick, mma4},
  laser2/.style   = { ultra thick,blue}
  photon/.style = {decorate, decoration={snake}, draw=red}
}

\begin{document}

\title{Photon entanglement entropy as a probe of many-body correlations and fluctuations}

\author{Hao Li}
\affiliation{Department of Chemistry, University of Houston, Houston, TX 77204}

\author{Andrei Piryatinski}
\affiliation{Theoretical Division, Los Alamos National Lab, Los Alamos, NM 87545}

\author{Ajay~Ram~Srimath~Kandada}
\affiliation{School of Chemistry \& Biochemistry and School of Physics, Georgia Institute of Technology, 901 Atlantic Drive, Atlanta, Georgia 30332}
\affiliation{Center for Nano Science and Technology @Polimi, Istituto Italiano di Tecnologia, via Giovanni Pascoli 70/3, 20133 Milano, Italy}

\author{Carlos Silva}
\affiliation{School of Chemistry \& Biochemistry and School of Physics, Georgia Institute of Technology, 901 Atlantic Drive, Atlanta, Georgia 30332}

\author{Eric R. Bittner}
\affiliation{Department of Chemistry \& Department of Physics, University of Houston, Houston, TX 77204}
\email{bittner@uh.edu}

\date{\today}

\begin{abstract}
Recent theoretical and experiments have
explored the use of entangled photons as a spectroscopic
probe of material systems.
We develop here a theoretical description for entropy
production in the scattering of an entangled biphoton
state within an optical cavity.  We  develop this
using perturbation theory by expanding the biphoton scattering matrix in terms of single-photon terms in
which we introduce the photon-photon interaction via a
complex coupling constant, $\xi$.   We show that the
von Neumann entropy provides a succinct measure of this
interaction.  We then
develop a microscopic model and show that
in the limit of fast fluctuations, the
entanglement entropy vanishes
whereas in the limit the coupling is homogeneous broadened, the entanglement entropy depends upon the
magnitude of the fluctuations
and reaches a maximum.

\end{abstract}

\maketitle

\section{Introduction}

Dynamical light scattering provides a
sensitive probe of correlations and
fluctuations in material systems.  The basic
theory and first applications of this technique harken back to experiments by Tyndall on aerosols in the 1860's and theoretical work by
Rayleigh.\cite{Tyndall2,Tyndall:1863aa,Tyndall1,Rayleigh:1871a,Rayleigh:1871b,Rayleigh:1871c} In fact, Rayleigh showed that light scattering from
density fluctuations in the atmosphere give rise to the blue color of the sky.\cite{Rayleigh:1899a}
The classic text by Berne and Pecorra helped to  establish the modern theory of
dynamical light scattering as an important
probe of chemical physical processes.\cite{Berne:DLS}

Experiments using entangled photon pairs as probes of
material systems have opened a new arena for both linear and non-linear
spectroscopy since quantum entangled photons facilitate a direct probe of  many-body
correlations.
 \cite{Ladd2010,Lemos2014,RevModPhys.74.145,Sewell2017,Zou1991a,PhysRevA.79.033832,Schlawin2013,Li_QS&T:2018}
With this in mind, we develop here a
theoretical approach that
connects the
resultant entropic change within a biphoton state to the matter-mediated
coupling between the photons within the sample.
Our theory develops from a perturbative expansion of the biphoton amplitude
in which the single-photon terms are coupled order-by-order via a
a complex {\em entanglement parameter}, $\xi$, which we take
as a measure of the photon-photon coupling mediated by the medium.
We also develop a microscopic model for photon-photon entanglement
a mediated by cross-correlated spectral fluctuations and relate this
to the von Neumann entropy of the outgoing biphoton quantum state.
We show that in the limit of rapid fluctuations and motional narrowing
destroy entanglement whereas in the limit of homogeneous broadening (gaussian noise),
fluctuations produce entangled states with a maximum entropy determined
by the spectral width.

\begin{figure*}
\label{Fig:ExSetup}
\includegraphics{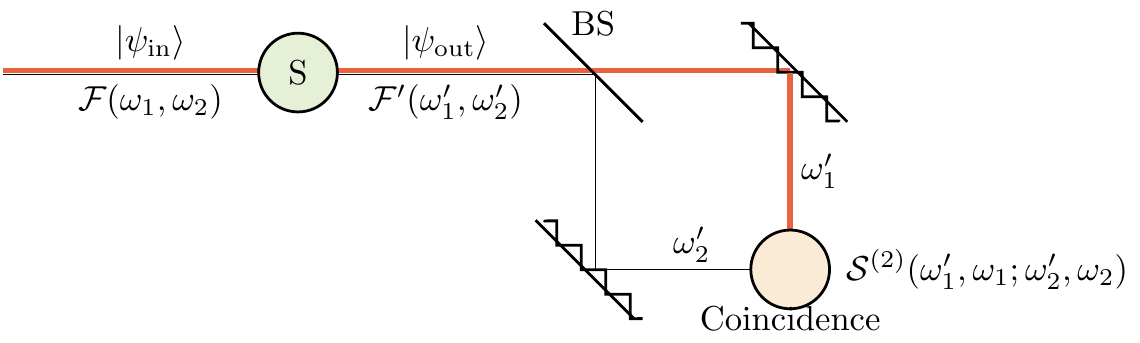}
\caption{Spectrally-resolved Hanbury Brown and Twiss experimental setup.
Following the beam-splitter (BS), the two photon frequencies are resolved before
coincidence detection.
}
\label{HBT}
\end{figure*}

\begin{widetext}

\section{Diagrammatic representation of two-photon scattering amplitude}
As illustrated in Fig.~\ref{Fig:ExSetup}, we consider here the entanglement produced by the interaction of an initial biphoton input state\cite{Loudon_QOBook:2000}
\begin{eqnarray}
\label{psi-in-Four}
|\psi_{\text{in}}\rangle &= &\iint d\omega_{1} d\omega_{2}{\cal F}(\omega_{1},\omega_{2})\hat B^{\dagger}(\omega_1)\hat B^{\dagger}(\omega_2)|0\rangle
\end{eqnarray}
with a sample $S$ to produce a biphoton output state
\begin{eqnarray}
\label{psi-out-Four}
|\psi_{\text {out}}\rangle &=& \iint d\omega'_{1} d\omega'_{2}\iint d\omega_{1} d\omega_{2}
\\
&~&
 {\cal S}^{(2)}(\omega_{1},\omega_{2};\omega'_{1},\omega'_{2}){\cal F}(\omega_{1},\omega_{2})\nonumber  \hat B^{\dagger}(\omega_{1})\hat B^{\dagger}(\omega_{2})|0\rangle
\end{eqnarray}
Here, we denote photon creation operators $\hat B^\dag (\omega_1)$ and $\hat B^\dag(\omega_2)$ acting  on the photon vacuum state $|0\rangle$, and the scattering amplitude $ {\cal S}^{(2)}$ reflecting the photons  interaction with the sample.  We shall leave the exact representation of the scattering amplitude undefined at the moment; however, we assume throughout that interaction with the sample placed before the beam-splitter affects the two-photon entanglement as the result of many-body interactions occurring when two photons interact with each other via the medium.  In principle, initial input ${\cal F}(\omega_{1},\omega_{2})$ may be separable into  single photon terms ${\cal F}(\omega_{1},\omega_{2}) = f_1(\omega_1)f_2(\omega_2)$  but we shall assume that the output amplitude is not separable and corresponds to an entangled photon pair.

Following interaction with the sample, the light passes through a symmetrical beam splitter (BS).  For co-linear photon beams, this produces the mapping\cite{Loudon_QOBook:2000,Li_QS&T:2018}
\begin{eqnarray}
\label{BtoA}
\hat B^{\dagger}(\omega'_{1})\hat B^{\dagger}(\omega'_{2})\mapsto \frac{1}{2} [\hat A^{\dagger}_{1}(\omega'_{1})+i\hat A^{\dagger}_{2}(\omega'_{1})][\hat A^{\dagger}_{1}(\omega'_{2})+i\hat A^{\dagger}_{2}(\omega'_{2})],
 \end{eqnarray}
where $A^{\dagger}_{1}$  ($A^{\dagger}_{2}$) creates a photon state in channel 1 (channel 2) after the beam splitter resulting in the post-beam splitter state given by
\begin{eqnarray}
|\psi_{BS}\rangle
&=&\frac{1}{2}\left( |\psi_{1} \rangle  - |\psi_{2} \rangle  + |\psi_{c}\rangle \right)
\end{eqnarray}
where the kets
$|\psi_{1} \rangle$  and $|\psi_{2}\rangle$ correspond to the cases where both photons are in either channel, and 
\begin{eqnarray}
\label{psi-co}
|\psi_{c}\rangle &=& \frac{i}{2} \iint d\omega'_{1} d\omega'_{2} \iint d\omega_{1} d\omega_{2} {\cal S}^{(2)}(\omega_{1},\omega_{2};\omega'_{1},\omega'_{2}) {\cal F}(\omega_{1},\omega_{2})[\hat{A}_{1}^\dagger(\omega_{1})\hat{A}_{2}^\dagger(\omega_{2}) + \hat{A}_{2}^\dagger(\omega_{1})\hat{A}_{1}^\dagger(\omega_{2})] |0\rangle \\
&=& \frac{i}{2} \iint d\omega'_{1} d\omega'_{2} \iint d\omega_{1} d\omega_{2} [{\cal S}^{(2)}(\omega_{1},\omega_{2};\omega'_{1},\omega'_{2}) {\cal F}(\omega_{1},\omega_{2}) + {\cal S}^{(2)}(\omega_{2},\omega_{1};\omega'_{2},\omega'_{1}) {\cal F}(\omega_{2},\omega_{1})] |\omega_{1}\omega_{2}\rangle
\end{eqnarray}
corresponds to the case where single photons are in both channels leading to the possibility of detecting the  coincidence counts.
From this we compute the coincidence probability as
\begin{eqnarray}
\label{Pc}
P_{c} = \langle \psi_{c} | \psi_{c} \rangle = &\frac{1}{4}&\iint d\omega_{1}d\omega_{2} \iint d\omega_{1}'d\omega_{2}'\{|{\cal S}^{(2)}(\omega_{1},\omega_{2};\omega_{1}',\omega_{2}')|^{2} |{\cal F}(\omega_{1},\omega_{2})|^{2} + |{\cal S}^{(2)}(\omega_{2},\omega_{1};\omega_{2}',\omega_{1}')|^{2} |{\cal F}(\omega_{2},\omega_{1})|^{2} \\
&+& 2\mathrm{Re}[{\cal S}^{(2)}(\omega_{1},\omega_{2};\omega_{1}',\omega_{2}') {\cal F}(\omega_{1},\omega_{2}) {\cal S}^{(2) *}(\omega_{2},\omega_{1};\omega_{2}',\omega_{1}') {\cal F}^{*}(\omega_{2},\omega_{1})]\}
\end{eqnarray}
which we take to be the integrated number of coincident photon pairs counted per unit time. For the experimental set-up sketched in Fig.~\ref{Fig:ExSetup},
symmetry dictates that ${\cal S}^{(2)}(\omega_{1},\omega_{2};\omega_{1}',\omega_{2}')={\cal S}^{(2)}(\omega_{2},\omega_{1};\omega_{2}',\omega_{1}')$ and ${\cal F}(\omega_{1},\omega_{2})={\cal F}(\omega_{2},\omega_{1})$, leading
to a counting rate of
\begin{equation}
P_{c} = \iint d\omega_{1}d\omega_{2} \iint d\omega_{1}'d\omega_{2}'|{\cal F}(\omega_{1},\omega_{2})|^{2} |{\cal S}^{(2)}(\omega_{1},\omega_{2};\omega'_{1},\omega'_{2})|^{2}.
\end{equation}
Hence, by measuring the spectral or polarization resolved coincidences,
one can reconstruct the biphoton scattering probability.
We next develop
 a relation between the scattered biphoton amplitude ${\cal S}^{(2)}$ and
the spectral response of the system.

\subsection{Diagrammatic Expansion of Scattering Amplitude}

 In general, we can write the elastic scattering
of a single photon through a resonant medium in the form
\begin{eqnarray}
{\cal S}^{(1)}(\omega,\omega') =
\exp({\cal A}(\omega))\delta(\omega-\omega')
\end{eqnarray}
where
\begin{eqnarray}
{\cal A}(\omega) = -\frac{ib}{(\omega_{o} - \omega) + i \gamma}
\end{eqnarray}
is the Fourier transform of the free induction decay
\begin{eqnarray}
{\cal A}(t) = -\sqrt{2\pi} b e^{-\gamma t} e^{i\omega_{o}t}
\end{eqnarray}
 for $t>0$ of an oscillator with frequency $\omega_o$ and dephasing time $1/\gamma$. $b = \alpha L\gamma/2$ where $\alpha L$ is the optical thickness and $\alpha$ is a Bouger coefficient.
Consequently, if two independent (non-entangled) photons are scattered from the resonant medium
we anticipate a scattering amplitude of
\begin{eqnarray}
{\cal S}^{(2)}(\omega_1,\omega_2;\omega'_1,\omega'_2) = {\cal S}^{(1)}(\omega_1)  {\cal S}^{(1)}(\omega_2)
\delta(\omega_1-\omega_1')\delta(\omega_2-\omega_2')
\end{eqnarray}
In this case, two independent photons are transmitted
without any interaction leading to them being entangled.

Suppose, however, that interactions leading to
entanglement are weak such that we can write the
two photon scattering amplitude as perturbation expansion
of the form
\begin{eqnarray}
{\cal S}^{(2)}&=& {\cal S}^{(2)}_{o} + {\cal S}_{o}^{(2)} {\cal V} {\cal S}^{(2)} = {\cal S}^{(2)}_{o} + {\cal S}_{o}^{(2)} {\cal V} {\cal S}_{o}^{(2)} + {\cal S}_{o}^{(2)} {\cal V} {\cal S}_{o}^{(2)} {\cal V} {\cal S}_{o}^{(2)} + \cdots \nonumber
\end{eqnarray}
whereby ${\cal S}^{(2)}_{o} = {\cal S}^{(1)}{\cal S}^{(1)}$ is separable
into single photon terms and  ${\cal V} $
mediates the interaction between photon pairs via the
resonant cavity.
This suggests the following diagrammatic
expansion
\begin{eqnarray}
\begin{tikzpicture}
\draw (0.1,-1)--(0.1,1);
\draw (-0.1,-1)--(-0.1,1);
\draw[thick,circle,fill=mma3!20](0,0) circle [radius=0.4];
\node at (0,0){${\cal S}^{(2)}$};
\node at (1,0){$=$};
\node at (0,-1.2) {$(\omega_{1},\omega_{2})$};
\node at (0,1.2) {$(\omega'_{1},\omega'_{2})$};
\draw (1.8,-1)--(1.8,1);
\draw (2.2,-1)--(2.2,1);
\node at (2.2,-1.2) {$\omega_{2}$};
\node at (1.8,-1.2) {$\omega_{1}$};
\node at (2.2,1.2) {$\omega'_{2}$};
\node at (1.8,1.2) {$\omega'_{1}$};
\node at (3,0) {$+$};
\draw (3.7,-1)--(3.7,1);
\draw (4.3,-1)--(4.3,1);
\draw [decorate,decoration={coil,segment length=4pt}] (3.7,0)--(4.3,0);
\node at (3.7,-1.2) {$\omega_{1}$};
\node at (4.3,-1.2) {$\omega_{2}$};
\node at (3.7,1.2) {$\omega'_{1}$};
\node at (4.3,1.2) {$\omega'_{2}$};
\node at (5,0) {$+$};
\draw (5.7,-1)--(5.7,1);
\draw (6.3,-1)--(6.3,1);
\draw [decorate,decoration={coil,segment length=4pt}] (5.7,0.3)--(6.3,0.3);
\draw [decorate,decoration={coil,segment length=4pt}] (5.7,-0.3)--(6.3,-0.3);
\node at (5.7,-1.2) {$\omega_{1}$};
\node at (6.3,-1.2) {$\omega_{2}$};
\node at (5.7,1.2) {$\omega'_{1}$};
\node at (6.3,1.2) {$\omega'_{2}$};
\node at (7,0) {$+ \cdots $};
\end{tikzpicture}
\end{eqnarray}
where solid lines are ${\cal S}^{(1)}$ propagations and springs denote the interaction.
Suppose we write that ${\cal V}(\omega)$ contributes a phase-shift of the form
\begin{eqnarray}{\cal V}(\omega) = \xi e^{i\phi}\delta(\omega)\label{Vint}\end{eqnarray}
but does not create a frequency shift.
Then only the term at $\omega = 0$ will contribute (so that $\omega_{1} = \omega_{1}'$ and $\omega_{2} = \omega_{2}'$)
\begin{eqnarray}
 {\cal S}_{1}^{(2)} = \xi e^{i\phi}
   {\cal S}^{(1)}(\omega_{1}){\cal S}^{(1)}(\omega_{2})
   {\cal S}^{(1)}(\omega_{1}){\cal S}^{(1)}(\omega_{2})
   \end{eqnarray}
Iterating this,
\begin{eqnarray}
{\cal S}_{n}^{(2)} &=& {\cal V}^{n} \left({\cal S}_{o}^{(2)}\right)^{n+1} = \xi^{n} e^{in\phi}
\left({\cal S}^{(1)}(\omega_{1})\right)^{n+1}
\left({\cal S}^{(1)}(\omega_{2})\right)^{n+1}
\end{eqnarray}
Taking
\begin{eqnarray}
{\cal S}^{(1)}(\omega_{i}) = \exp\left[-\frac{i b}{(\omega_{o} -  \omega_{i}) +  i \gamma} \right] = z_{i} \label{eq:s1}
\end{eqnarray}
is a complex number determined by the input photon frequency.
Thus, the whole perturbation series becomes
\begin{eqnarray}
{\cal S}^{(2)} 
= z_{1}z_{2} \sum_{n=0}^{\infty}\left(\xi e^{i\phi} z_{1}z_{2} \right)^{n}
\end{eqnarray}
Setting $q = \xi e^{i\phi}z_{1}z_{2}$ and assuming $|q| < 1$
 then the series can be summed exactly
\begin{eqnarray}
{\cal S}^{(2)} = \frac{z_{1}z_{2}}{1-q} 
\end{eqnarray}
Writing this in terms of the ${\cal S}^{(1)}$ functions, we obtain
\begin{eqnarray}
{\cal S}^{(2)}(\omega_{1},\omega_{2};\omega_{1},\omega_{2})
= \frac{{\cal S}^{(1)}(\omega_{1}){\cal S}^{(1)}(\omega_{2})}{1-\xi {\cal S}^{(1)}(\omega_{1}){\cal S}^{(1)}(\omega_{2})e^{i\phi}}. \label{eq:s2}
\end{eqnarray}
We shall refer to $\xi$ as the {\em entanglement parameter}.
When $\xi =  0$,
\begin{eqnarray}
{\cal S}^{(2)} (\omega_{1},\omega_{2}) =\exp[ {\cal A}^{(1)}(\omega_{1})]\exp[ {\cal A}^{(1)}(\omega_{2})] = {\cal S}^{(1)} (\omega_{1}) {\cal S}^{(1)} (\omega_{2})
\end{eqnarray}
is separable in terms of the individual photon amplitudes.

\subsection{Averaging over Phase}
In principle, the phase $\phi$ introduced in Eq.~\ref{Vint} depends upon the
microscopic details of the system, such as the relative orientation
of the atomic or molecular scattering sites within the sample, and
may be simply be a random quantity.
Averaging over phase, we write
\begin{eqnarray}
\overline{\cal S}^{(2)}(\omega_{1},\omega_{2};\omega_{1},\omega_{2})
=
\left \langle
\frac{{\cal S}^{(1)}(\omega_{1}){\cal S}^{(1)}(\omega_{2})}{1-\xi {\cal S}^{(1)}(\omega_{1}){\cal S}^{(1)}(\omega_{2})e^{i\phi}}
\right\rangle.
\end{eqnarray}
Writing this again as a geometric series
\begin{eqnarray}
\overline{\cal S}^{(2)}(\omega_{1},\omega_{2};\omega_{1},\omega_{2})
=
{\cal S}^{(1)} (\omega_{1}) {\cal S}^{(1)} (\omega_{2})
\sum_n
(\xi {\cal S}^{(1)} (\omega_{1}) {\cal S}^{(1)} (\omega_{2}))^n
\left\langle
e^{in \phi}
\right\rangle.
\end{eqnarray}
Suppose that the phase $\phi$ is uniform over $[0,2\pi)$, then
$$
\left\langle
e^{in \phi}
\right\rangle  = \int_0^{2 \pi} \frac{e^{i n \phi}}{2 \pi} d \phi = \delta_{n0}.
$$
In this case, the relative phase  is completely randomized and the
biphoton amplitude collapses {\em exactly} into the product of two single photon terms.
\begin{eqnarray}
\overline{\cal S}^{(2)}(\omega_{1},\omega_{2};\omega_{1},\omega_{2})
\to
{\cal S}^{(1)} (\omega_{1}) {\cal S}^{(1)} (\omega_{2})
\end{eqnarray}

On the other hand, suppose the phase is normally distributed about a central value, which we can take to be zero,
{\em i.e. } $\overline \phi = 0$ and $\overline{\phi^2} = \sigma^2$.
Here, the average over $\phi$ can be cast as
$$
\left\langle
e^{in \phi}
\right\rangle  = \left(
1- \frac{n^2 \overline{\phi^2}}{2!} +   \frac{n^4 \overline{\phi^4}}{4!} - \cdots \right)
$$
Writing this in terms of the second moment
\begin{eqnarray}
\left\langle
e^{in \phi}
\right\rangle  =
\sum_{k=0}^\infty (-1)^k (n \sigma)^{2k} \frac{(2k-1)!!}{(2k)!}  = \sum_{k=0}^\infty (-1)^k \frac{(n \sigma)^{2k} }{2^k k!} = e^{-(n\sigma)^2/2}.
\end{eqnarray}
This gives
\begin{eqnarray}
\overline{\cal S}^{(2)}(\omega_{1},\omega_{2};\omega_{1},\omega_{2})
=
\sum_{n=0}^\infty
\xi^n e^{-n^2 \sigma^2/2} ( {\cal S}^{(1)} (\omega_{1}) {\cal S}^{(1)} (\omega_{2}))^{n+1}.
\label{gaussian-noise}
\end{eqnarray}
Plots for  this are shown in Fig.~\ref{fig3}a for the case of a model system with
spectral parameters  for Eq.~(\ref{eq:s1}): $\omega_0 = \pi$, $b_o = 1$ and $\gamma = 2$.
In each case, we assume an input of
two photon Fock-state giving an output
biphoton amplitude that is correlated along $\omega_1 = \omega_2$.
For the case of Gaussian noise,
the final state is not necessarily separable into the product of two
functions and the resulting state is
correlated in frequency, as shown in Fig.~\ref{fig3} for
various choices of spectral parameters.
The von Neumann entropy, $S_\psi$ (c.f. Sec.~\ref{SecIIIb})
gives a useful means of
quantifying the entanglement of these states.
Schmidt decomposition of Eq.~\ref{gaussian-noise} for a
given value of $\sigma$ gives Fig.~\ref{fig3}b where
we have plotted the von Neumann entropy
in terms of increasing interaction and in terms of
increasing Gaussian noise.
Generally, increasing $\xi$ leads to an increase in
entanglement for a given amount of noise $\sigma$ as shown in Fig.~\ref{fig3}b.

\begin{figure*}
\subfigure[]{\includegraphics[width = 0.54\columnwidth]{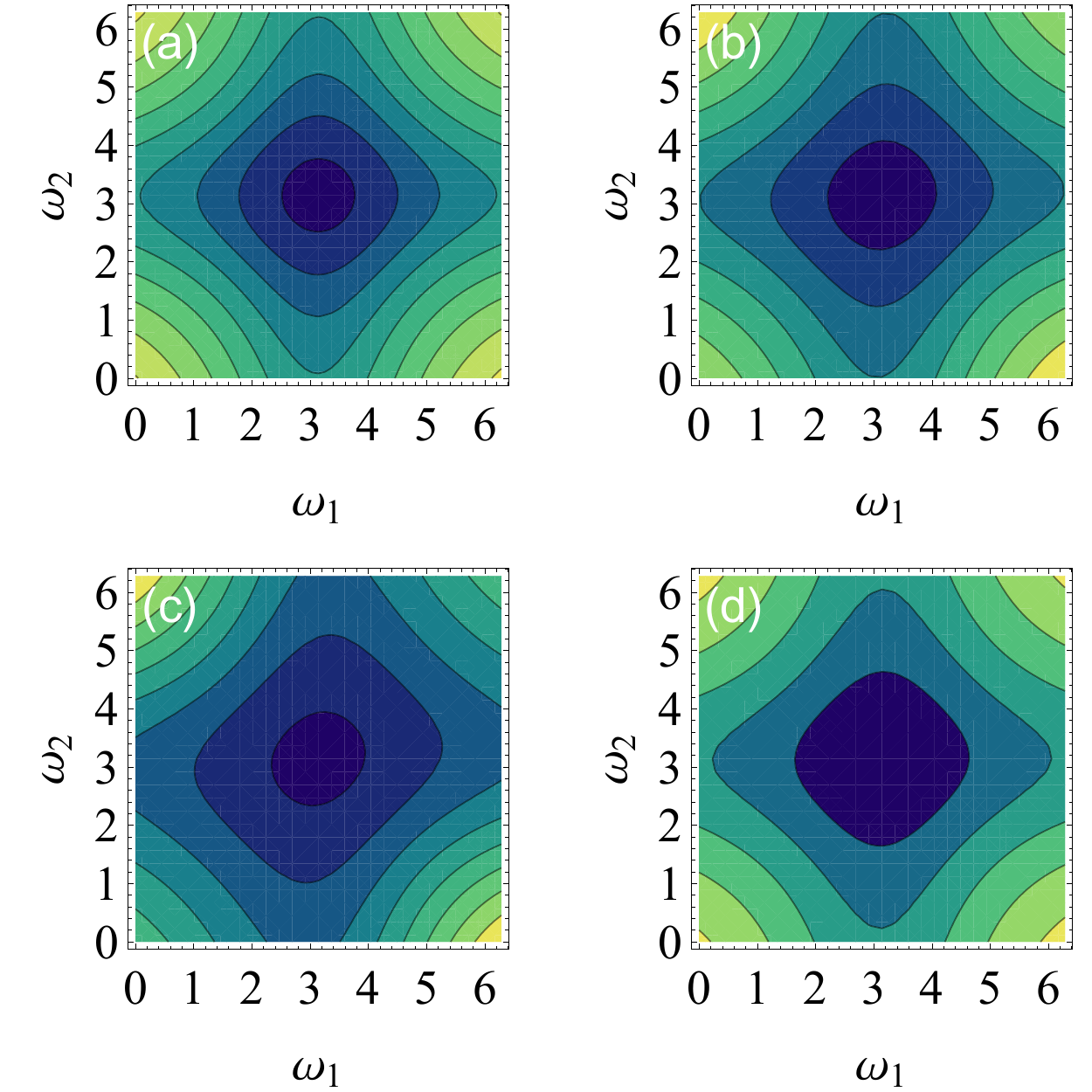}}
\subfigure[]{\includegraphics[width =0.45\columnwidth]{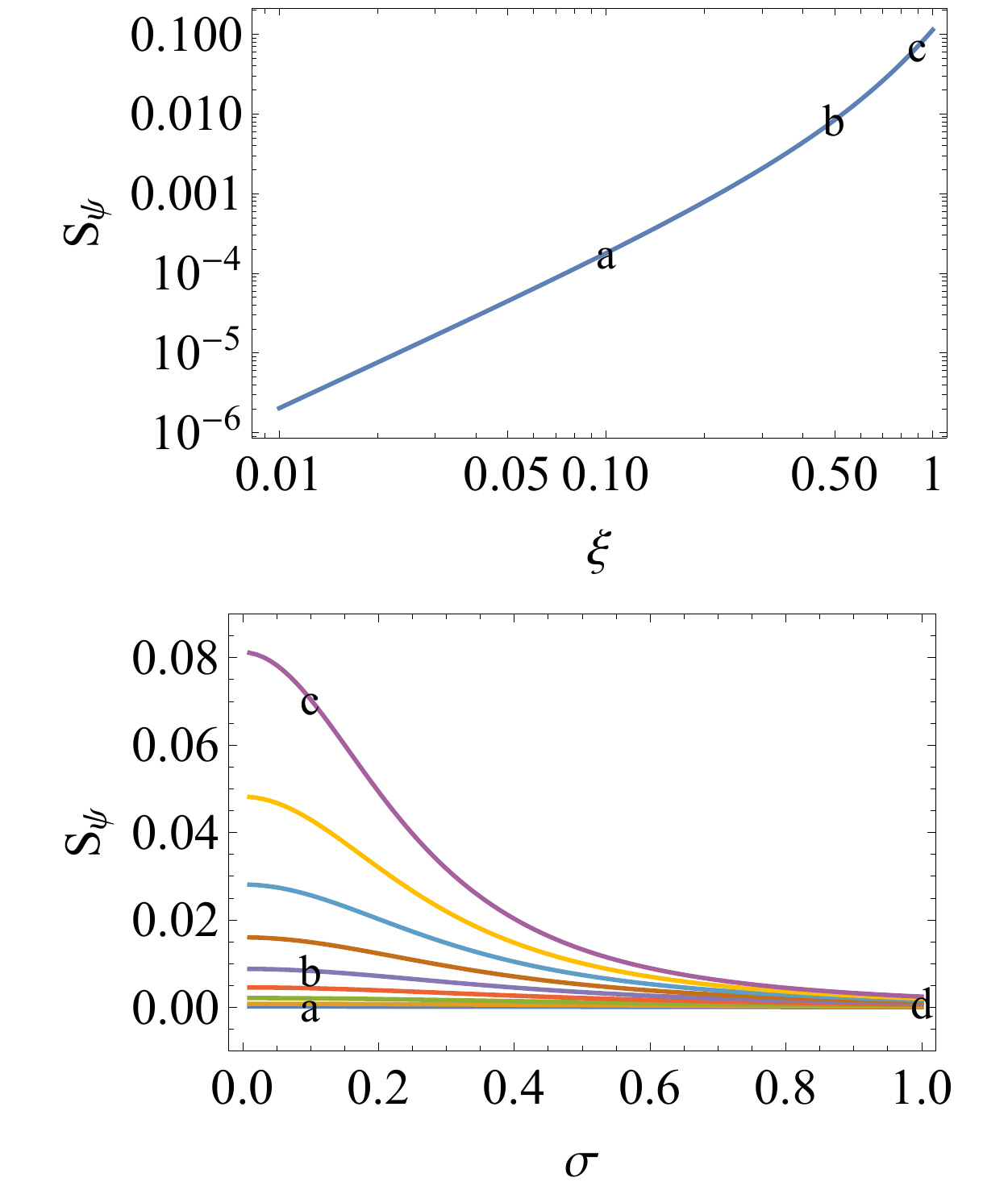}}
\caption{
(Left) Absolute values of two-photon scattering functions for squeezed biphotons with Gaussian noise interactions:
(a) $\xi = 0.1$, $\sigma = 0.1$;
(b) $\xi = 0.5$, $\sigma = 0.1$;
(c) $\xi = 0.9$, $\sigma = 0.1$;
(d) $\xi = 0.5$, $\sigma = 1.0$.
In each case we take parameters for ${\cal S}^{(1)}$ in
Eq.~(\ref{eq:s1}) as: $\omega_0 = \pi$, $b_o = 1$ and $\gamma = 2$.
(Right,top)
Entanglement entropy (Eq.~(\ref{eq:entropy})) vs. coupling $\xi$ for case of gaussian noise.  Points a,b, and c correspond to
specific biphoton states shown to the left.
(Right,bottom) Entanglement entropy (Eq.~(\ref{eq:entropy}))  vs. noise strength ($\sigma$).  Points a to d correspond to
specific biphoton states shown to the left.
}
\label{fig3}
\end{figure*}

\end{widetext}

\section{Stochastic model for two-photon scattering amplitude and entanglement entropy}

We now dig deeper and develop a fully microscopic model
for entropy production in biphoton scattering.
In time-domain the input state on the left boundary of the cavity (Fig.~\ref{Fig:ExSetup}) can be represented as
\begin{eqnarray}
\label{psi-in-t}
|\psi_{\text{in}}\rangle =\iint dt_1 dt_2{\cal F}(t_1,t_2)\hat b_{1,\text{in}}^\dag(t_1)\hat b_{2,\text{in}}^\dag(t_2)|0\rangle,
\end{eqnarray}
where the cavity input photon operator\cite{Li_QS&T:2018}
\begin{eqnarray}
\label{bin-def}
\hat b^\dag_{j,\text{in}}(t)=-\int \frac{d\omega}{\sqrt{2\pi}}\hat B_j^\dag(\omega)e^{i\omega t}
\end{eqnarray}
is the Fourier transform of the external photon mode operator $\hat B_j^\dag(\omega)$ entering Eq.~(\ref{psi-in-Four}). Noteworthy that substitution of Eq.~(\ref{bin-def}) into Eq.~(\ref{psi-in-t}) and subsequent integration over the time variables exactly results in the frequency domain representation given by Eq.~(\ref{psi-in-Four}) where ${\cal F}(\omega_1,\omega_2)$ is identified as the Fourier transform of ${\cal F}(t_1,t_2)$. Following, the input output formalism,\cite{GardinerOC:1984,CollettPRA:1984} we identify the boundary condition on the left boundary of the cavity with the mode leakage rate $\kappa_j$ as $\hat b_{j,\text{in}}^\dag(t)+\hat b_{j,\text{r}}^\dag(t)=\sqrt{\kappa_j}\hat b_j^\dag(t)$.  This connects the input photon operator defined above, the {\it reflected} (output) photon operator $\hat b_{j,\text{r}}^\dag(t)$, and the cavity photon operator $b_j^\dag(t)$. The reflected mode is not measured in experiment and will not be considered below. However, the cavity mode operator contains information on the interactions within the cavity and will be determined below.

Time domain cavity output operators can be expressed in terms of the photon operators outside the cavity as \cite{Li_QS&T:2018}
\begin{eqnarray}
\label{bout-def}
\hat b^\dag_{j,\text{out}}(t)=\int \frac{d\omega}{\sqrt{2\pi}}\hat B_j^\dag(\omega)e^{i\omega (t-t_f)},
\end{eqnarray}
representing interaction free photon modes $\hat B_j^\dag(\omega)$ outside the cavity propagated back in time from $t_f$ to actual measurement time $t$. Evaluation of the coincidence probability requires partitioning of $\hat B_j^\dag(\omega)$ operator according to Eq.~(\ref{BtoA}). Thus, the output state needs to be represented in terms of these operators as $|\psi_{\text{out}}\rangle=\iint d\omega'_1 d\omega'_2 {\cal F}(\omega'_1,\omega'_2) \hat B_1^\dag(\omega'_1)\hat B_2^\dag(\omega'_2)|0\rangle$ which according Eq.~(\ref{bout-def}) translates to
\begin{eqnarray}
\label{psi-out-t}
|\psi_{\text{out}}\rangle &=&\iint d\omega_1 d\omega_2
{\cal F}(\omega_1,\omega_2)
\iint dt_1 dt_2e^{-i\omega_1(t_1-t_f)}
\\\nonumber&\times&
e^{-i\omega_2 (t_2-t_f)}
\langle\hat b_{1,\text{out}}^\dag(t_1)\hat b_{2,\text{out}}^\dag(t_2)\rangle|0\rangle,\;\;\;\;
\end{eqnarray}
where and angle brackets describe the average of the output operators over the material induced cavity mode fluctuations. Similar to the input mode we establish the following boundary condition on the cavity right boundary $\hat b_{j,\text{out}}^\dag(t)=\sqrt{\kappa_j}\hat b_j^\dag(t)$. This condition assumes the same cavity leakage, $\kappa_j$, and according to the setup geometry in Fig.~\ref{Fig:ExSetup}, no input photons on the right. Taking into account this boundary condition,  evaluation of the output state using Eq.~(\ref{psi-out-t}) reduces to the evaluation of the time domain correlation function of the cavity mode operators that will directly result in the evaluation of the desired scattering amplitude.

\subsection{Stochastic Model for Cavity Photon Scattering}

We adopt the following stochastic Hamiltonian to describe cavity photon modes $\hat{b}_j^\dag$ coupled to the input and output biphoton states   
\begin{eqnarray}
\label{Hph-stch}
	\hat H_{\text{ph}}=\hbar\sum\limits_{j=1,2} \left(\omega_j+\delta\omega_j\right)\hat b_j^\dag \hat b_j.
\end{eqnarray}
Here, $\delta\omega_j=\delta\omega_j(t)$ is the time-dependent photon frequency fluctuations of each mode.  Appendix~\ref{Appx:Hstch} provides an example connecting such a generic Hamiltonian with a microscopic Hamiltonian describing photon wave packet scattering by fluctuations of delocalized polariton modes confined within a cavity.

Applying input output formalism to the cavity modes described by the Hamiltonian~(\ref{Hph-stch}), one obtains the quantum Langevin equation\cite{Gardiner:1985,Li_QS&T:2018}
\begin{eqnarray}
\label{eq-mov-bdag}
	\frac{\partial}{\partial t}\hat b_j^\dag(t) = i \left(\tilde\omega_j+\delta\omega_j\right)\hat b_j^\dag(t)
			+\sqrt{\kappa_j}\hat b_{j, \text{in}}^\dag(t).
\end{eqnarray}
with $\tilde\omega_j=\omega_j+i\kappa_j/2$. Assuming that the material fluctuation dynamics occurs on the timescale faster that the cavity leakage one can formally integrate Eq.~(\ref{eq-mov-bdag}) that results in
\begin{eqnarray}
\label{bdag-opsol}
	\hat b_{j,\text{out}}^\dag(t) &=&\sqrt{\kappa_j}\int_0^t dt' e^{i\tilde\omega_j(t-t')}
\\\nonumber&\times&
\exp_+\left[i\int\limits_{t'}^t d\tau \delta\omega_j(\tau)\right]   \hat b_{j,\text{in}}^\dag(t'),
\end{eqnarray}
with $\exp_+[\dots]$ being positive time ordered exponential. Here, we also used the boundary condition for the cavity right boundary to express the cavity mode operator in terms of cavity output mode.

According to Eq.~(\ref{bdag-opsol}), the output single and two-photon operators averaged over the cavity fluctuations can be represented as
\begin{eqnarray}
\label{1ph-state-t-daf}
&~&\langle \hat b_{j,\text{out}}^\dag(t)\rangle  =\int_0^\infty dt' {\cal S}_j^{(1)}(t,t')f_j(t')\hat b_{\text{in},j}^\dag(t'),
\\\label{2ph-state-t-daf}
&~&\langle\hat b_{1,\text{out}}^\dag(t_1)\hat b_{2,\text{out}}^\dag(t_2)\rangle =\int_0^\infty dt_1'\int_0^\infty dt_2'
\\\nonumber &~&\hspace{20pt}
{\cal S}^{(2)}(t_1t_2,t_1't_2') {\cal F}(t_1',t_2')\hat b_{1,\text{in}}^\dag(t_1') \hat b_{2,\text{in}}^\dag(t_2'),
\end{eqnarray}
where we explicitly assumed ${\cal F}(\omega_{1},\omega_{2}) = f_1(\omega_1)f_2(\omega_2)$ for a single photon propagation. The single- and two-photon scattering amplitudes entering Eqs.~(\ref{1ph-state-t-daf}) and (\ref{2ph-state-t-daf}) read
\begin{eqnarray}
\label{St-1ph-def}
&~&{\cal S}_j^{(1)}(t,t') = \theta(t-t') e^{i\tilde\omega_j(t-t')}
\\\nonumber&~&\hspace{40pt}\times
\left\langle \exp_+\left[i\int\limits_{t'}^t d\tau \delta\omega_j(\tau)\right] \right \rangle,
\\\label{St-2ph-def}
&~&{\cal S}^{(2)}(t_1t_2,t_1't_2') =\theta(t_1-t_1')\theta(t_2-t_2')
\\\nonumber&~&\hspace{60pt}\times
e^{i\tilde\omega_1 (t_1-t_1')+i\tilde\omega_2(t_2-t_2')}
\\\nonumber&~&\times 	
\left\langle 	\exp_+\left[i\int\limits_{t_1'}^{t_1} d\tau \delta\hat\omega_1(\tau)\right]
				\exp_+\left[i\int\limits_{t_2'}^{t_2} d\tau \delta\hat\omega_2(\tau)\right]
\right \rangle.
\end{eqnarray}
Here, $\theta(t)$ is the Heaviside theta-function and angle brackets indicate average over the frequency fluctuations. Substitution of Eq.~(\ref{2ph-state-t-daf}) and (\ref{St-2ph-def}) into Eq.~(\ref{psi-out-t}) provides an expression for the output biphoton state in terms of the scattering amplitude and biphoton input states allowing for the evaluation of the coincidence probability.

Diagrammatic techniques can be developed for single- and two-photon scattering amplitudes via power series expansion of the exponentials in Eqs.~(\ref{St-1ph-def}) and (\ref{St-2ph-def}). Instead, we adopt a second cumulant approximation setting all odd point correlation functions in the expansion to zero, and partition the rest in to various products of two-point correlation functions. Summation of the resulting power series gives rise to the following representation of the single- and two-photon scattering amplitudes\cite{kubo,Mukamel:1995}
\begin{eqnarray}
\label{St-1ph-cum}
&~&{\cal S}_j^{(1)}(t,t')= e^{i\tilde\omega_j(t-t')-g_j(t,t')},
\\\label{St-2ph-cum}
&~&{\cal S}^{(2)}(t_1t_2,t_1't_2')=e^{i\tilde\omega_1(t_1-t_1')+i\tilde\omega_2 (t_2-t_2)}
\\\nonumber&~& \hspace{60pt}\times e^{-g_1(t_1,t_1')-g_2(t_2,t_2')-g_{12}(t_1t_2,t_1't_2')},
\end{eqnarray}
respectively. Accordingly, the two-photon scattering amplitude can be factorized as
\begin{eqnarray}
\label{St-2ph-K}
{\cal S}^{(2)}(t_1t_2,t_1't_2')&=&{\cal S}_1^{(1)}(t_1,t_1'){\cal S}_1^{(1)}(t_2,t_2')
\\\nonumber&\times&{\cal K}^{(2)}(t_1t_2,t_1't_2'),
\end{eqnarray}
where the irreducible part
\begin{eqnarray}
\label{K12-2ph}
	{\cal K}^{(2)}(t_1t_2,t_1't_2')=e^{-g_{12}(t_1t_2,t_1't_2')},
\end{eqnarray}
is introduced.

The second cumulant function $g_j$ 
(Eqs.~(\ref{St-1ph-cum})) depends on the $j$-th frequency autocorrelation function as
\begin{eqnarray}
\label{gj}
g_j(t,t') = \int\limits_{t'}^td\tau_1\int\limits_{t'}^{\tau_1}d\tau_2
	\left\langle \delta\omega_j(\tau_1) \delta\omega_j(\tau_2) \right\rangle,
\end{eqnarray}
and gives rise to the photon dephasing. Noteworthy, the integration over $d\tau_1$ and $d\tau_2$ is time-ordered insuring causality for a single-photon propagation. The  second cumulant function entering the irreducible part (Eq.~(\ref{St-2ph-cum})) is
\begin{eqnarray}
\label{g12}
g_{12}(t_1t_2,t_1't_2') = \int\limits_{t_1'}^{t_1}d\tau_1\int\limits_{t_2'}^{t_2}d\tau_2
	\left\langle \delta\omega_1(\tau_1) \delta\omega_2(\tau_2) \right\rangle.\;\;\;\;\;\;
\end{eqnarray}
This accounts for the cross-correlations between different photon modes and as we show below affects the photon pair entanglement.  In contrast to Eq.~(\ref{gj}), here integration over $d\tau_1$ and $d\tau_2$ lacks time ordering indicating that the photon cross-correlations are not casual. If the cross-correlation function is zero then ${\cal K}_{12}=1$ and the two-photon scattering amplitude factorized to a product of single-photon ones.

For further analysis, we adopt Kubo-Anderson model which is often used in spectroscopic line shape analysis.\cite{kubo} This model treats fluctuations as commuting random variable, whose time evolution is a Gaussian stochastic process making second cumulant expansion exact. Following this approach, we set
\begin{eqnarray}
\label{Corr-Gauss-j}
\left\langle \delta\omega_j(\tau_1) \delta\omega_j(\tau_2) \right\rangle &=& \sigma_j^2e^{-|\tau_1-\tau_2|/\bar\tau_j},
\\\label{Corr-Gauss-12}
\left\langle \delta\omega_{1}(\tau_1) \delta\omega_2(\tau_2) \right\rangle&=&\sigma_{12}^2
					e^{-|\tau_1-\tau_2|/\bar\tau_{12}},
\end{eqnarray}
where $\sigma_j^2=\left\langle \delta\omega_j^2(0)\right\rangle$ and $\sigma_{12}^2=\left\langle \delta\omega_{1}(0)\delta\omega_{2}(0)\right\rangle$ ($\bar\tau_j$ and $\bar\tau_{12}$) being single-mode and cross-mode variances (correlation times), respectively.

The representation of the single photon amplitudes is not essential for the analysis below and
the details of the derivation of the
 second cumulant functions in Eqs.~(\ref{gj}) and (\ref{g12}) for the correlation functions given in Eqs.~(\ref{Corr-Gauss-j}) and (\ref{Corr-Gauss-12})
are given in Appendix~\ref{Appx:g-calc}. We discuss the limiting cases here.

In the limit of inhomogeneous broadening where  $\sigma_{12}\bar\tau_{12}\ll 1$,
only four time-ordered contributions (denoted by $g_{12}^+$) survive:
\begin{eqnarray}
\label{Kt-in-1}
e^{-g^+_{12}(t_1t_2t_1't_1')} &=& e^{-\gamma_{12}(t_2 - t_1')},
\\\label{Kt-in-2}
e^{-g^+_{12}(t_1t_2t_2't_1')} &=& e^{-\gamma_{12}(t_2 - t_2')},
\\\label{Kt-in-3}
e^{-g^+_{12}(t_2t_1t_2't_1')} &=& e^{-\gamma_{12}(t_1 - t_2')},
\\\label{Kt-in-4}
e^{-g^+_{12}(t_2t_1t_1't_2')} &=& e^{-\gamma_{12}(t_1 - t_2')}.
\end{eqnarray}

 In the  limit of slow modulation for the mode cross-correlation, {\em i.e.}  $\sigma_{12}\bar\tau_{12}\gg 1$ only
 terms that are  quadratic in time contribute to the second cumulant function.
 In optics, this is referred to as the homogeneous broadening limit. In this case,
 the irreducible part of the two-photon scattering amplitude acquires a Gaussian form
\begin{eqnarray}
\label{Kt-2ph-hom}
	{\cal K}^{(2)}(t_1t_2,t_1't_2')=e^{-\sigma_{12}^2(t_1-t_1')(t_2-t_2')}.
\end{eqnarray}

\subsection{Entanglement Entropy Analysis}
\label{SecIIIb}

Using the irreducible part of two-photon scattering amplitude introduced in Eq.~(\ref{K12-2ph}), we can compute the von Neumann entropy
$S = - Tr[\rho \ln \rho]$ for the scattered biphoton state.
Whereas above, we computed this in the
frequency domain, $S$ is invariant under
unitary transformations, including the Fourier transform,
so we should be able to evaluate the entropy directly from the
time-correlation functions.
This can by accomplished by performing a
Schmidt decomposition of the scattering amplitude in
Eq.~(\ref{St-2ph-K}) into separable components.  Since this is
a product of  separable and non-separable  terms, we only need to
decompose the irreducible part (Eq.~(\ref{K12-2ph})) involving $g_{12}$,
\begin{eqnarray}
\label{Sdecomp}
e^{-g_{12}(t_1,t_2,t_1',t_2')} &=&\sum_k r_k \phi_k(t_1-t_1') \psi_k(t_2-t_2')
\end{eqnarray}
where the functions  $\{ \phi_k(t_1-t_1')\}$ and $\{ \psi_k(t_2-t_2') \}$
form an orthonormal basis of Schmidt modes and $r_k$ are
the mode weights.
The mode weights provide a useful way to quantify the entanglement between photons.

If we write $\lambda_{k} = r_{k}/\sqrt{B}$ as the set of normalized Schmidt coefficients
such that
\begin{eqnarray}
\sum_{k}\lambda_{k}^{2}= 1\end{eqnarray}
we can write the von Neumann entropy as
\begin{eqnarray}
S= -\sum_{k}\lambda_{k}^{2} \ln(\lambda_{k}^{2}) \label{eq:entropy}
\end{eqnarray}
If the state is a pure state, then the entropy is exactly zero  and one and only one of the
$\lambda_k^2= 1$, the rest are exactly equal to zero.  That is to say that
the biphoton state is separable.
Moreover, $S = \ln N$ where $N$ is the dimensionality of the Hilbert-space spanned by the
basis functions.   In other words, increasing $S$ implies that  more and more pairs of
Schmidt basis functions are needed to reconstruct the original function.

In the first case, where $\sigma_{12}\bar\tau_{12} \ll 1$ (which would be the
limit of motional narrowing), the exponent of the cross-correlation function is separable
in terms of the times (Eqs.~(\ref{Kt-in-1})-(\ref{Kt-in-4})) and consequently, the entropy of the out-going state
is exactly equal to 0.  This makes sense since in this limit the cross-correlation
function depends only upon the intermediate two times in the time-ordering.
In other words, the only way for photon 1 to interact with photon 2 is if the
polarization created by the first persists long enough to influence the second photon.
Else, no additional entanglement can be produced.

\begin{figure}
\subfigure[]{\includegraphics[width=0.45\columnwidth]{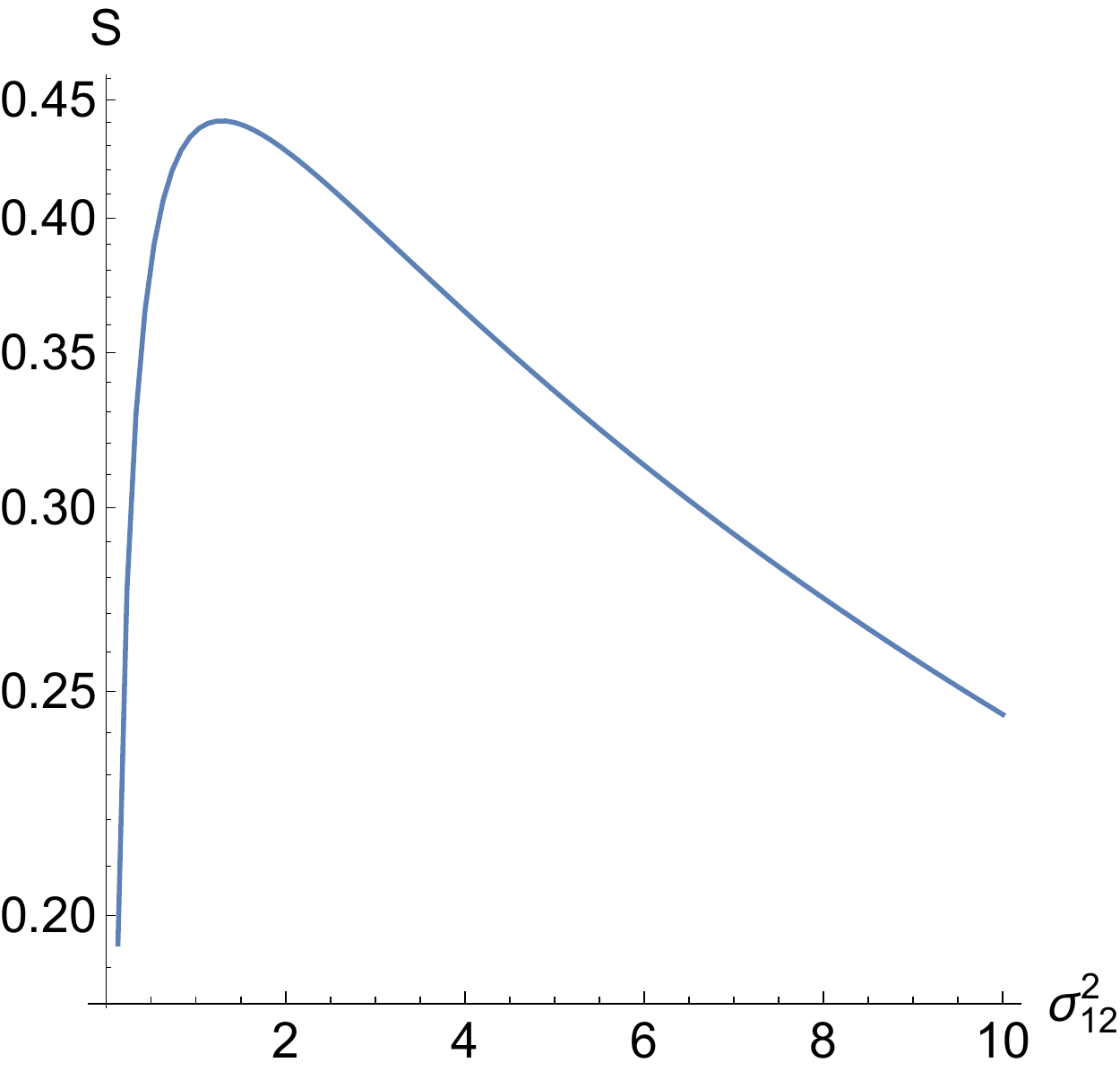}}
\subfigure[]{\includegraphics[width=0.45\columnwidth]{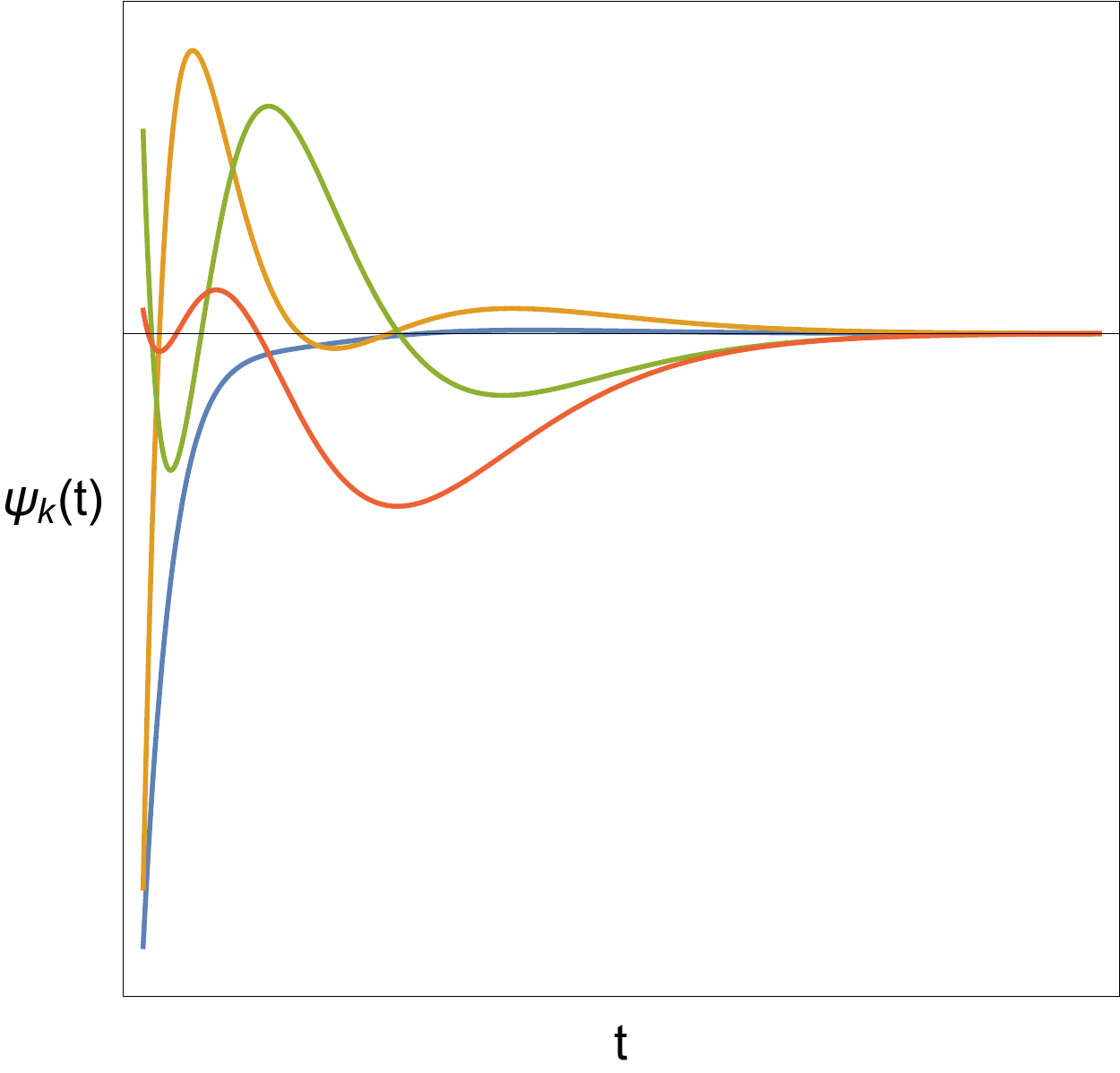}}
\caption{(a) Entropy vs. cross-correlation magnitude $\sigma_{12}$ in the case of
slow-modulation (Homogenous broadening limit).  (b) Schmidt modes for
maximally entangled state ($\sigma_{12}^2 = 1.33$).} \label{homo-ent}
\end{figure}

In the limit of slow modulation, the cross-correlation depends upon all 4 times (Eq.~(\ref{Kt-2ph-hom})) and can not be separable
into a pair of functions involving only $t_1-t_1'$ and $t_2-t_2'$. 
Here, we first expand Eq.~(\ref{Kt-2ph-hom}) as a sum products of Laguerre polynomials
$$
e^{-\sigma_{12}^2x y} = \sum_{nm} c_{nm} w^{1/2}(x) w^{1/2}(y) L_n(x) L_m(y)
$$
where $w(x) $ are the Gaussian quadrature weights
and then determined the Schmidt vectors and coefficients by diagonalizing the matrix $c.c^\dagger$.
\cite{Lamata}
Fig.~\ref{homo-ent}a shows the resulting entropy for this limit as a function of the
fluctuation strength $\sigma_{12}^2$.    Interestingly, this shows a maximum in the
entanglement for $\sigma_{12}^2 \approx 1.33$.   This can be understood
in the following way.      In the limit that $\sigma_{12}^2$ is small, fluctuations
are simply too weak to generate entanglement.    On the other hand,
large fluctuations will lead to decoherence and collapse any entanglement
that may be presents.   The maximum, then falls in the  limit of  being neither
too soft nor too hard. \footnote{We suggest this state be termed the ``Goldilocks State''.}
Fig.~\ref{homo-ent}b shows the Schmidt basis functions for the maximal entropy
case where  $\sigma_{12}^2$.

An interesting case arises when
$$
{\cal K}^{(2)} = a\delta(t-t').
$$
Since one can write  the delta-function
as a resolution of the identity in terms of orthogonal polynomials $O_n(x)$, \cite{Lamata}
\begin{eqnarray}
\delta(t-t') = \lim_{N\to \infty}\sum_{k=0}^N
O_k(t)O_k(t'),
\end{eqnarray}
its normalized Schmidt coefficients are all equal to
$1/\sqrt{N}$.  This gives rise to the case of
maximal entanglement entropy since
\begin{eqnarray}
S =-\lim_{N
\to \infty} \sum_{k = 0}^N
\frac{1}{N}\log(1/N) \to \infty.
\end{eqnarray}
\section{Discussion}

We have presented here a model for the generation of
entanglement entropy for a biphoton Fock state
interacting with a material sample.  We assume that the
two-photon scattering matrix can be decomposed into
a series of single photon/photon interactions mediated by
coupling to a medium with coupling strength $\xi$.
In the limit that the scattering
produces a random phase shift, the entanglement collapses and the outgoing state is a single Fock state.
However, in the case of Gaussian noise, the
entanglement entropy increases with increasing coupling
producing squeezed states.
We also present a microscopic model for the photon-photon
coupling for the case of two photons passing through
an optical cavity.  Here, we again show that in the limit
of fast fluctuations and motional narrowing, the
entanglement entropy vanishes where as in the case
of slow modulation (homogeneous broadening) the entanglement entropy reaches a maximum value depending upon the magnitude of the
fluctuations.
Our analysis shows that two-photon entanglement
scattering provides
a direct and sensitive probe of
correlated fluctuations
within the  sample system.

\acknowledgements

The work at the University of Houston was funded in
part by the  National Science Foundation (CHE-1664971, MRI-1531814)
and the Robert A. Welch Foundation (E-1337).
AP acknowledges the support provided by Los Alamos National Laboratory Directed Research and Development (LDRD) Funds.
CS acknowledges support from the School of Chemistry \& Biochemistry and the College of Science of Georgia Tech.
ARSK acknowledges funding from  EU Horizon 2020 via Marie Sklodowska Curie Fellowship (Global) (Project No. 705874).

\section*{Author Contributions}

 ERB, CS, and AP conceived and developed the core ideas.
 ERB, HL, and AP developed and implemented the formalism.
 All authors participated in the discussion of the methods and results.
 All authors participated in the drafting
 and final editing of this manuscript.

\section*{Competing financial interests}

The authors declare no competing financial interests.

\appendix
\section{Photon mode scattering via cavity polariton fluctuations}
\label{Appx:Hstch}

In this Appendix adopted stochastic Hamiltonian~(\ref{Hph-stch}) is derived based on a simple model describing light scattering by the fluctuations of delocalized polariton modes confined in an optical cavity. For this situation, the photon Hamiltonian can be represented as a sum of two components
\begin{eqnarray}
\label{Hph-scat-def}
	\hat H_{\text{ph}}=\hat H_{\text o}+ \hat H_{\text{s}}.
\end{eqnarray}
Assuming that biphoton wavepacket is spatially confined within a cross section of area $A$ and propagates in the $z$-direction, the interaction free photon Hamiltonian in the continues mode representation reads\cite{Loudon-book}
\begin{eqnarray}
\label{Hph0}
	\hat H_{\text o}=\sum\limits_{j=1,2}\int d\omega_j \omega_j \hat b_{\omega_j}^\dag \hat b_{\omega_j}.
\end{eqnarray}
with index $j$ distinguishing the modes.

The photon scattering illustrated in Fig.~\ref{Fig:ExSetup} is described by the interaction Hamiltonian
\begin{eqnarray}
\label{Hph-scat-def}
	\hat H_{\text s}= - \iint\limits_A dxdy \int\limits_{-L/2}^{L/2}dz \sum\limits_{j=1,2}\hat\alpha_{jj}(\bm r)\hat E_j(z)^2,
\end{eqnarray}
where the integration $dS$ over the photon wave packet cross section ($x,y$-plane) is partitioned from the spatial integral in the propagation direction $z$. The cavity cross-section is assumed to be larger than $A$ and the length is denoted by $L$. The electric field operator for the photon modes of interest represented in terms of the mode creation and annihilation operators reads\cite{Loudon-book}
\begin{eqnarray}
\label{E-qntz}
	\hat E_j(z)&=&i\int d\omega_j \sqrt{\frac{\hbar\omega_j}{4\pi\varepsilon_o c A}}
	\left(b_{\omega_j}e^{i\omega_j z/c}-b_{\omega_j}^\dag e^{-i\omega_j z/c}\right),\;\;\;\;\;\;\;
\end{eqnarray}
where $c$ is speed of light and $\varepsilon_o$ is the vacuum permittivity.

Evaluation of the integrals in the scattering Hamiltonian~(\ref{Hph-scat-def}) requires a model for the sample polarizability operator $\hat\alpha(\bm r)$. Let us consider delocalized cavity polariton modes which we described by operator
\begin{eqnarray}
\label{zeta-def}
\hat\zeta_{\bm k_l}=\bar \zeta_{\bm k_l}+\delta\hat\zeta_{\bm k_l}(t).
\end{eqnarray}
where $\bm k_l$ denotes $s$-th polariton mode wave vector. $\bar \zeta_{\bm k_l}$ is a cavity polariton steady state prepared by a resonant external pumping and $\delta\hat\zeta_{\bm k_l}(t)$ is time-dependent mode fluctuation operator. Accordingly, the polarizability can be expanded up to the first order in the fluctuations
\begin{eqnarray}
\label{pol-flct-ex}
	\hat\alpha_{jj}&=&\alpha_{jj}(\bar\zeta_{\bm k_l})+\sum\limits_l\sum\limits_{\bm k_l}
		\frac{\partial\alpha_{jj}(\bar\zeta_{\bm k_l})}{\partial\bar\zeta_{\bm k_l}}\delta\hat\zeta_{\bm k_l}(t).
\end{eqnarray}
For further analysis the fluctuation operator is expanded in terms of polariton spatial Fourier components
\begin{eqnarray}
\label{fluct-pm}
	\delta\hat\zeta_{\bm k_l}(t)= \delta\hat\zeta^+_{\bm k_l}(t)e^{i\bm k_l\bm r}+\delta\hat\zeta_{\bm k_l}^-(t)e^{-i\bm k_l\bm r}.
\end{eqnarray}

Making substitution of the second term in Eq.~(\ref{pol-flct-ex}) along with Eq.~(\ref{fluct-pm})  into the scattering Hamiltonian~(\ref{Hph-scat-def}) where the electric field is introduced by Eq.~(\ref{E-qntz}), performing integration over the cavity volume, and further neglecting the terms describing simultaneous two-photon creation and annihilation processes, we obtain
\begin{eqnarray}
\label{Hph-scat-k}
\hat H_{\text s} &=& - \hbar \sum_l\sum\limits_{k_l}\sum\limits_{j=1,2}
	\int d\omega_{j} \kappa_{k_l}(\omega')
\\\nonumber&\times&	
	\left( \hat b_{\omega_{j}+c k_l}^\dag  \hat b_{\omega_j} \delta\hat\zeta^+_{k_l}
	+\hat b_{\omega_j}^\dag   \hat b_{\omega_j+ck_l} \delta\hat\zeta^-_{k_l} \right),
\end{eqnarray}
 with the coupling parameter
\begin{eqnarray}
\label{kappa-wkz}
	\kappa_{k_l}(\omega_j)&=&\frac{1}{\varepsilon_oL}\left.\frac{\partial\alpha_{jj}(\bar\zeta_{\bm k_l})}
	{\partial\bar\zeta_{\bm k_l}}\right|_{\bm k^\bot_l=0}\hspace{-20pt}\left[(\omega_j-ck_l)\omega_j\right]^{1/2}.
\end{eqnarray}
Since the photons propagate in $z$-direction, the total momentum conservation requires that the scattered photon momentum changes for the amount of $k_j$ which is the $z$-component of the total momentum $\bm k_j$. Accordingly, the transverse ($xy$-plane) component of the momentum $\bm k_j^\bot$ does not change resulting in the photon coupling to the $\Gamma$-point of transverse polariton band as indicated above by setting $\bm k^\bot_j=0$.

Taking into account that polariton modes have continuous dispersion relations $k_l=k(\omega_l)$ , we replace sum over $k_l$ in Eq.~(\ref{Hph-scat-k}) by the integral over $d\omega_l$. This results in
\begin{eqnarray}
\label{Hph-scat-w}
\hat H_{\text s} &=& - \hbar \sum\limits_{j=1,2}\sum_l\iint d\omega_{j} d\omega_l~\kappa(\omega_j\omega_l)
\\\nonumber &\times&	
	\left( \hat b_{\omega_{j}+\omega_l}^\dag  \hat b_{\omega_j} \delta\hat\zeta^+_{\omega_l}
	+\hat b_{\omega_j}^\dag   \hat b_{\omega_j+\omega_l} \delta\hat\zeta^-_{\omega_l} \right),
\end{eqnarray}
with the coupling parameter
\begin{eqnarray}
\label{kappa-w1w2}
	\kappa(\omega_j\omega_l)&=&\frac{1}{2\pi\varepsilon_o}\left.\frac{\partial\alpha_jj(\bar\zeta_{\bm k(\omega_l)})}
	{\partial\bar\zeta_{\bm k(\omega_l)}}\right|_{\bm k^\bot(\omega_l)=0}
\\\nonumber&\times&	
	\frac{\partial k(\omega_l)}{\partial \omega_l}
	\left[(\omega_j-\omega_l)\omega_j\right]^{1/2}.
\end{eqnarray}

The scattering Hamiltonian~(\ref{Hph-scat-w}) can be further simplified, provided the interaction occurs near the bottom of polariton modes, i.e. $\omega_l\sim 0$. In this case one can set
\begin{eqnarray}
\label{dalpha-loc}
	\left.\frac{\partial\alpha_{jj}(\bar\zeta_{\bm k(\omega_l)})}
	{\partial\bar\zeta_{\bm k(\omega_l)}}\right|_{\bm k^\bot(\omega_l)=0}
	=\bar\alpha_{jj}\delta(\omega_l).
\end{eqnarray}
with $\bar\alpha_{jj}$ being a coupling constant. Substitution of Eq.~(\ref{kappa-w1w2}) with Eq.~({dalpha-loc}) into Eq.~(\ref{Hph-scat-w}) recasts the latter to the form of stochastic Hamiltonian~(\ref{Hph-stch}) with the frequency fluctuation operator defined as 
\begin{eqnarray}
\label{w-fluct-def}
	\delta\omega_j =\frac{\bar\alpha_{jj}\omega_j}{2\pi\varepsilon_o}
	\sum_l\frac{\partial k(\omega_l)}{\partial \omega_l}\delta\hat\zeta_l.
\end{eqnarray}
where a shorthand notation $\delta\hat\zeta_l=\delta\hat\zeta_{\omega_l=0}$ is used.

\section{Evaluation of second cumulants for Gaussian stochastic process}
\label{Appx:g-calc}

In this Appendix, we derive explicit form of the single-mode and cross-mode cumulant functions using the correlation functions for stochastic Gaussian processes given in Eqs.~(\ref{Corr-Gauss-j}) and (\ref{Corr-Gauss-12}), respectively.

Evaluation of time ordered integral in Eq.~(\ref{gj}) with the correlation function given by Eq.~(\ref{Corr-Gauss-j}) results in a well known form of the second cumulant\cite{kubo}
\begin{eqnarray}
\label{gj-Kubo}
g_j(t-t') &=& (\sigma_j\bar\tau_j)^2
\\\nonumber&\times&
\left\{(t-t')/\bar\tau_j+e^{-(t-t')/\bar\tau_j}+1\right\}.\;\;\;\;\;\;
\end{eqnarray}
In the case of inhomogeneous broadening $\sigma\bar\tau_j\ll 1$ one gets
\begin{eqnarray}
\label{gj-Kubo-inh}
g_j(t-t') &=& \gamma_j(t-t'),
\end{eqnarray}
with $\gamma_j=\sigma^2\bar\tau_j$. In the case of homogeneous broadening $\sigma\bar\tau_j\gg 1$ one gets
\begin{eqnarray}
\label{gj-Kubo-hom}
g_j(t-t') &=& \sigma_j^2(t-t')^2.
\end{eqnarray}

Evaluation of the integrals in Eq.~(\ref{g12}) with the cross-correlation function in Eq.~(\ref{Corr-Gauss-12}) is not so straight forward and one needs to take into account various ordering of $t_1'$, $t_1$, $t_2'$ and $t_2$:

\begin{tikzpicture}
\draw [->] (0,0)--(5,0);
\node at (-0.5,0) {$A:$};
\node at (-.1,0) {$t$};
\node at (5.3,0) {$\infty$};
\draw [->,decorate,decoration={snake},red] (0.,1)--(.5,0);
\node at (.5,0.2) {$t_2'$};
\draw [->,decorate,decoration={snake},red] (1.5,0)--(2,1);
\node at (1.5,0.2) {$t_2$};
\draw [->,decorate,decoration={snake},red] (2.0,-1)--(2.5,0);
\node at (2.5,-0.2) {$t_1'$};
\draw [<-,decorate,decoration={snake},red] (4.0,-1)--(3.5,0);
\node at (3.5,-0.2) {$t_1$};
\end{tikzpicture}

\begin{tikzpicture}
\draw [->] (0,0)--(5,0);
\node at (-0.5,0) {$B:$};
\node at (-.1,0) {$t$};
\node at (5.3,0) {$\infty$};
\draw [->,decorate,decoration={snake},red] (0.,1)--(.5,0);
\node at (.5,0.2) {$t_2'$};
\draw [->,decorate,decoration={snake},red] (1.5,0)--(1,-1);
\node at (2.5,0.2) {$t_2$};
\draw [->,decorate,decoration={snake},red] (3.0,1)--(2.5,0);
\node at (1.5,-0.2) {$t_1'$};
\draw [<-,decorate,decoration={snake},red] (4.0,-1)--(3.5,0);
\node at (3.5,-0.2) {$t_1$};
\end{tikzpicture}

\begin{tikzpicture}
\draw [->] (0,0)--(5,0);
\node at (-0.5,0) {$C:$};
\node at (-.1,0) {$t$};
\node at (5.3,0) {$\infty$};
\draw [->,decorate,decoration={snake},red] (0.,1)--(.5,0);
\node at (.5,0.2) {$t_2'$};
\draw [->,decorate,decoration={snake},red] (1.5,0)--(1,-1);
\node at (3.5,0.2) {$t_2$};
\draw [<-,decorate,decoration={snake},red] (3.0,-1)--(2.5,0);
\node at (1.5,-0.2) {$t_1'$};
\draw [->,decorate,decoration={snake},red] (4.0,1)--(3.5,0);
\node at (2.5,-0.2) {$t_1$};
\end{tikzpicture}

\noindent
with three more corresponding to swapping index $1$ and $2$ (but not the primes).
\begin{eqnarray}
\label{g12-Kubo-1}
g^+_{12}(t_1t_1't_2t_2') &=& (\sigma_{12}\bar\tau_{12})^2\left\{
\right.\\\nonumber &~&\left.
	e^{-(t_1-t_2)/\bar\tau_{12}}-e^{-(t_1-t_2')/\bar\tau_{12}}
\right.\\\nonumber &+&\left.
	e^{-(t_1'-t_2)/\bar\tau_{12}}-e^{-(t_1'-t_2')/\bar\tau_{12}}		
\right\},\;\;\;\;
\end{eqnarray}
\begin{eqnarray}
\label{g12-Kubo-2}
g^+_{12}(t_1t_2t_1't_2') &=& (\sigma_{12}\bar\tau_{12})^2\left\{2(t_2-t_1')/\bar\tau_{12}
\right.\\\nonumber &-&\left.
	e^{-(t_1-t_2)/\bar\tau_{12}}+e^{-(t_1-t_2')/\bar\tau_{12}}
\right.\\\nonumber &-&\left.
	e^{-(t_1'-t_2')/\bar\tau_{12}}+e^{-(t_2-t_1')/\bar\tau_{12}}		
\right\},\;\;\;\;
\end{eqnarray}
\begin{eqnarray}
\label{g12-Kubo-3}
g_{12}^+(t_1 t_2t_2't_1') &=& (\sigma_{12}\bar\tau_{12})^2\left\{2(t_2-t_2')/\bar\tau_{12}
\right.\\\nonumber &-&\left.
	e^{-(t_2'-t_1')/\bar\tau_{12}}+e^{-(t_2-t_1')/\bar\tau_{12}}
\right.\\\nonumber &-&\left.
	e^{-(t_1-t_2)/\bar\tau_{12}}+e^{-(t_1-t_2')/\bar\tau_{12}}		
\right\},\;\;\;\;
\end{eqnarray}
and three more
 \begin{eqnarray}
 \label{g12-Kubo-4}
 g^+_{12}(t_2t_1t_1't_2') &=& (\sigma_{12}\bar\tau_{12})^2\left\{2(t_1-t_1')/\bar\tau_{12}
 \right.\\\nonumber &-&\left.
 	e^{-(t_1'-t_2')/\bar\tau_{12}}+e^{-(t_1-t_2')/\bar\tau_{12}}
 \right.\\\nonumber &-&\left.
 	e^{-(t_2-t_1)/\bar\tau_{12}}+e^{-(t_2-t_1')/\bar\tau_{12}}		
 \right\},\;\;\;\;
 \end{eqnarray}
 \begin{eqnarray}
 \label{g12-Kubo-5}
 g^+_{12}(t_2t_1t_2't_1') &=& (\sigma_{12}\bar\tau_{12})^2\left\{2(t_1-t_2')/\bar\tau_{12}
 \right.\\\nonumber &-&\left.
 	e^{-(t_2-t_1)/\bar\tau_{12}}+e^{-(t_2-t_1')/\bar\tau_{12}}
 \right.\\\nonumber &-&\left.
 	e^{-(t_2'-t_1')/\bar\tau_{12}}+e^{-(t_1-t_2')/\bar\tau_{12}}		
 \right\},\;\;\;\;
 \end{eqnarray}
 \begin{eqnarray}
 \label{g12-Kubo-6}
 g^+_{12}(t_2t_2't_1t_1') &=& (\sigma_{12}\bar\tau_{12})^2\left\{
 \right.\\\nonumber &~&\left.
 	e^{-(t_2-t_1)/\bar\tau_{12}}-e^{-(t_2-t_1')/\bar\tau_{12}}
 \right.\\\nonumber &+&\left.
 	e^{-(t_2'-t_1)/\bar\tau_{12}}-e^{-(t_2'-t_1')/\bar\tau_{12}}		
 \right\},\;\;\;\;
 \end{eqnarray}
with time indices 1 and 2 swapped. Note, our notation is such that in $g_{12}^+(t_at_bt_ct_d)$, $ t_a > t_b> t_c > t_d$.

In the case of {\em inhomogeneous} broadening $\sigma_{12}\bar\tau_{12}\ll 1$, $g^+_{12}(t_1t_1't_2t_2') =0 $, $g^+_{12}(t_2t_2't_1t_1') =0 $, and the rest of time-ordered cumulants simplify to the form
\begin{eqnarray}
\label{g12-inh-1}
g^+_{12}(t_1t_2t_1't_2') &=& \gamma_{12}(t_2-t_1'),
\\\label{g12-inh-2}
g_{12}^+(t_1 t_2t_2't_1') &=& \gamma_{12}(t_2-t_2'),
\\\label{g12-inh-3}
g^+_{12}(t_2 t_1 t_2't_1') &=& \gamma_{12}(t_1-t_2'),
\\\label{g12-inh-4}
g_{12}^+(t_2 t_1t_1't_2') &=& \gamma_{12}(t_1-t_1'),
\end{eqnarray}
with $\gamma_{12}=\sigma_{12}^2\bar\tau_{12}$.

In the case of {\em homogeneous} broadening $\sigma_{12}\bar\tau_{12}\gg 1$, Eqs.~(\ref{g12-Kubo-1})-(\ref{g12-Kubo-6}) simplify to the following expression
\begin{eqnarray}
\label{A:g12-Kubo-en}
g_{12}(t-t') &=& \sigma_{12}^2(t_1-t_1')(t_2-t_2'),
\end{eqnarray}
which holds for all initial time permutations used in Eqs.~(\ref{g12-Kubo-1})-(\ref{g12-Kubo-6}) .


%

\end{document}